\documentclass[a4paper]{jpconf}
\usepackage[dvips]{graphicx}
\usepackage{amssymb}
\begin{document}

\title{\vspace{-3cm} \\Leaky-box approximation to the fractional diffusion model of cosmic rays}
\author{V. V. Uchaikin, R. T. Sibatov, V. V. Saenko}
\address{Ulyanovsk State University, 432017, Leo Tolstoy str., 42, Ulyanovsk, Russia}
\ead{vuchaikin@gmail.com}

\begin{abstract}

Two models of anomalous diffusion of cosmic ray in the leaky-box approximation \cite{Ptuskin2009} are compared: one of them is based on the decoupled time-space L\'evy flights and the other on fractional walks with a finite free motion velocity. Distributions of first passage time and paths are computed and evolution of diffusion packets to equilibrium state is shown. Calculations demonstrate essential difference between the two models: the coupled scheme gives more realistic results.

\end{abstract}

\section{Introduction}

We continue investigation of problems arising in the fractional model of cosmic rays propagation in the Galaxy. Recall the situation. The first version of this model was proposed in \cite{Lagutin:NuclPhysB:2001}. It was based on the three-dimensional L\'{e}vy-Feldheim flight process in an infinite homogeneous medium obeying the equation with a fractional power of Laplacian
$$
\left[\frac{\partial}{\partial t}+D(-\Delta)^{\alpha/2}\right]G(\mathbf{r},t)=\delta(\mathbf{r})\delta(t),
$$
containing the ``diffusion coefficient'' $D$, which depends on the energy $E$ and  remains constant in the process of motion, and the fractional Laplacian given by its Fourier transform
$$
\int e^{i\mathbf{k}\mathbf{r}}(-\triangle)^{\alpha/2}f(\mathbf{r})d\mathbf{r}=
|\mathbf{k}|^{\alpha}
\int e^{i\mathbf{k}\mathbf{r}}f(\mathbf{r})d\mathbf{r}\equiv|\mathbf{k}|^{\alpha}\widetilde{f}
(\mathbf{k}),\quad 0<\alpha\leq2.
$$
This equation and its solution expressed through the isotropic L\'{e}vy-Feldheim distribution~$\Psi^{(\alpha)}(\mathbf{x})$
$$
G(\mathbf{r},t)=(Dt)^{-3/\alpha}\Psi^{(\alpha)}(\mathbf{r}(Dt)^{-3/\alpha})
$$
were known to that time (see \cite{Uchaikin1999}). The fractional power of the Laplacian has been explained by a fractal (friable) large scale structure of the Galaxy which causes an enhanced kind of diffusion (superdiffusion). The form of such a diffusion packet is described by the isotropic Levy-Feldheim distribution, and its width grows with time proportionally to $t^{1/\alpha}$. When $\alpha=2$, the medium becomes homogeneous and anomalous diffusion reduces to the normal one with Gaussian profile and the width $\propto t^{1/2}$. As a result, we saw that this solution on the assumption of power type of energy dependence of diffusivity ($D\propto E^\delta$) and source spectrum ($S\propto E^{-p}$) reveals an effect similar to the observed ``knee'' in primary spectra. As noted at the end of the cited work, the best fit of experimental data for H, He, CNO, Ne-Si and Fe-group were observed at $\alpha=5/3\approx 1.67,\ \delta =0.25$ and $p=2,9$. We will refer to this model as the LU-model.

\section{Lagutin-Tyumentsev model}

Later, the model was modified by inserting a fractional time-derivative instead of the first-order one \cite{Lagutin:NIMPS:2003} and lowering the orders $\alpha$ of the fractional Laplacian to $0.3$ \cite{Lagutin:BulASU:2004}:
$$
\left[\ _0\textsf{D}_t^\beta+D(-\Delta)^{\alpha/2}\right]G(\mathbf{r},t)=\delta(\mathbf{r})\delta_\beta(t),\quad \delta_\beta(t)= t^{-\beta}/\Gamma(1-\beta),\quad \alpha=0.3,\quad \beta=0.8.
$$
For the sake of convenience, we will refer to this modification as the LT-model.

The following case for such a choice was given. The value $\beta=0.8$ was taken from the work \cite{Cadavid1999} devoted  to investigation how photospheric convective motions transport magnetic flux elements. Experimental data exhibit  subdiffusion behaviour of solar magnetic elements. Observations of solar magnetic bright points analyzed in the cited work led to conclusion that the waiting time distribution density follows $t^{-\beta-1},\quad \beta=0.61\pm0.09$, during interval 0.3-22 minutes and then rapidly damps. Thus, we do not see here any reasons for application of the power law with $\beta=0.8$ to description of propagation of cosmic rays through interstellar medium: these phenomena are quite different by space-time scales and even by nature. Moreover, the power-law behavior is observed at small times only and disappears in the long-time region. This corresponds  to the value $\beta=1$.

The spatial exponent $\alpha$ was changed firstly from  1.67 to 1.00 in \cite{Erlykin2003c}. The authors wrote that ``a comparison of simulated characteristics
with experiment indicates that the fractal structure
of ISM with the parameter $\alpha=1$ (Kraichnan
spectrum of magnetic irregularities) gives local
cosmic ray characteristics which are closest to the
experiment''. However, as far back as in 1999 \cite{Korobko1999}, see also \cite{Uchaikin2004}, was found, that the fractal dimension $d_F$ of a medium does not coincide  with the exponent $\alpha$ characterizing the free path distribution in this medium. This conclusion was supported by analytical calculations performed in \cite{Isliker2003}. They showed that $\alpha$ depends not only on $d_F$  but also on size of scattering objects (say, magnetic clouds).  When it was recognized, Lagutin, Raikin and Tyumentsev \cite{Lagutin:IzvASU:2004a} repeated the calculations and determined the spatial exponent $\alpha=0.3$ from the linearized relation
$$
\alpha=2-d_F
$$
with $d_F=1.7$ \cite{Lagutin:IzvASU:2004a}. The latter number for the fractal dimension of the interstellar medium can be considered as a conventional value (see, for example, \cite{Combes1999}). Link between $\alpha$ and $d_F$ is valid only in the limit of point fractal (\cite{Lagutin:IzvASU:2004a}). Approximately, this formula could be used in the case if the interstellar magnetic field form small islands located at large distances of each other, but really it is not the case. Interstellar magnetic clouds affecting the cosmic ray transport have various sizes and may be close to each other. Thus, expecting shorter free paths, one should take essentially larger values of $\alpha$. Looking at Figure~3 of the article \cite{Lagutin:IzvASU:2004a}, one can see an ambiguous $\alpha(d_F)$ dependence: the $\alpha$ is determined not only by the $d_F$ but also by ratio size/distance for inhomogeneities. When this ratio grows, the fractal becomes less transparent and the free path pdf falls more rapidly. We refer to the work \cite{Ketabi2009}, where the numerical simulation with the Erlykin-Wolfendale model gave $\alpha=1.6\div 1.9$ which, in the authors opinion, ``is expected result, implying a fractal structure for interstellar medium''. The choice is also supported by recent article \cite{Kermani2011}. Using the known cosmic ray spectrum and radial gradient in the vicinity of the solar system to define an energy density and comparing with the modeling results shoved that the best fit for the value of $\alpha$ is about 1.65 (possible fits range from 1.6-1.9, but not acceptable fit is found for $\alpha=2$, which would correspond to conventional diffusion). Our initial value $\alpha=1.67$ belongs to this region.

\begin{figure}[thb]
\centering
\includegraphics[width=1\textwidth]{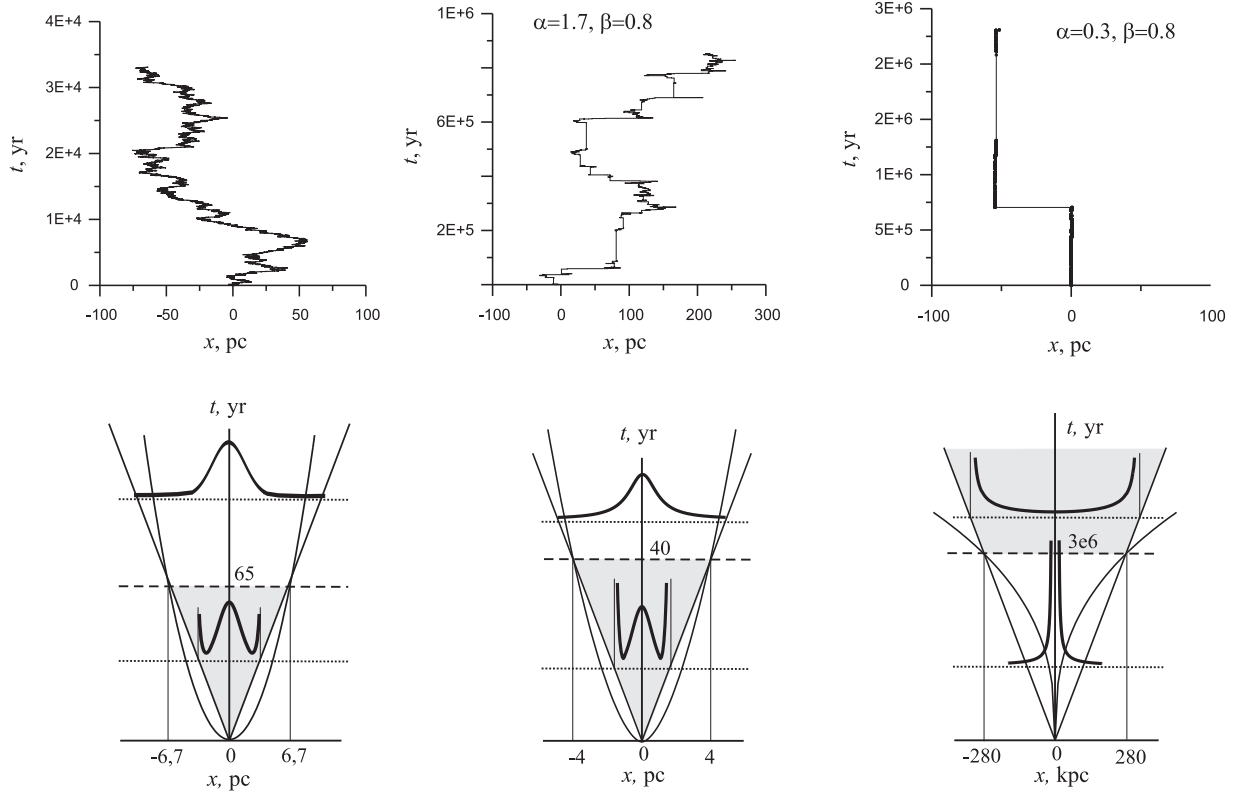}
\caption{Examples of time-position trajectories for three models: normal diffusion (path length $l=1$~pc), anomalous diffusion ($\alpha=1.67, \beta=0.8$,  ${D_0=2.4\cdot10^{-3}}$~$\mathrm{pc}^{1.7}/\mathrm{year}^{0.8}$, $D=D_0 R^{0.27}$, $E=10^6$~GeV), and LT model
  ($\alpha=0.3, \beta=0.8$, ${D_0=4\cdot10^{-6}}$~$\mathrm{pc}^{0.3}/\mathrm{year}^{0.8}$, $D=D_0 R^{0.27}$, $E=10^6$~GeV). The lower panels present
  schematically the diffusion packet spreading laws, shadowed regions correspond to regions with determining influence of speed finiteness.}\label{traj}
\end{figure}

\section{Bounded anomalous diffusion model}

In our works \cite{Uchaikin:JETPL:2010, Uchaikin2011, Uchaikin2011b} we show that random trajectories related to LT-model contains anomalously long rectilinear parts comparable with size of the Galaxy disc itself, and assumption on instantaneous flights to such distances looks to say the least of it unphysical (see Figure~\ref{traj}). The exit from this situation lies in using the fractional material derivative operator as it described in the above-sited our works: this operator takes into account that cosmic ray particles propagate through interstellar medium with a finite speed. the fractional material operator used in our bounded anomalous diffusion model \cite{Uchaikin2011a}, where the following equation
$$
A_\alpha\left\langle\ _0{\mathcal {D}}_t^\alpha\right\rangle G(\mathbf{r},t)=S_v(\mathbf{r},t),\quad 0<\alpha<1;
$$
$$
\left[\frac{1}{v}\frac{\partial}{\partial t}+A_\alpha\left\langle_0{\mathcal {D}}_t^\alpha\right\rangle\right] G(\mathbf{r},t)=S_v(\mathbf{r},t),\quad 1<\alpha<2;
$$
for cosmic ray propagation was represented. Here, $G(\mathbf{r},t)$ is the propagator,
$S_v(\mathbf{r},t)$ is a source function, the angle brackets
denote averaging over directions $\mathbf{\Omega}$ of propagation, the operator
$$
_0{\mathcal {D}}_t^\alpha G(\mathbf{r},t)=\left(\frac{1}{v}\frac{\partial}{\partial
t}+\mathbf{\Omega}\nabla\right)^\alpha G(\mathbf{r},t)
$$
is the fractional generalization of the material derivative. When $\alpha\in(1,2)$, the equation reduces in the long time asymptotic region to the Levy-flight diffusion equation
$$
\frac{\partial G}{\partial t}=-D_v(-\triangle)^{\alpha/2}G(\mathbf{r},t),\quad G(\mathbf{r},0)=\delta(\mathbf{r}),
$$
with diffusivity \cite{Zolotarev1999}
$$
D_v=\frac{D_\infty}{1+w/v},
$$
where $w$ stands for he mean path covered by L\'{e}vy-jumps per
unit time and $v$ is the free motion of particles. The solution of
Eq.~(2) for an unbounded fractal medium is expressed through the
isotropic L\'{e}vy-Feldheim distribution
$$
G(\mathbf{r},t)=(D_vt)^{-3/\alpha}\Psi^{(\alpha)}((D_vt)^{-1/\alpha}\mathbf{r}),
\quad 1<\alpha<2.
$$

\section{Fractional Laplacian in a bounded domain}

In this work, we consider another aspect of cosmic ray propagation, provoked by the long-distant parts, namely the influence of boundaries on the fractional Laplacian. Indeed, the true Laplacian is a local operator, having the same form independently of presence or absence of boundaries. However, the fractional Laplacian is a non-local operator and for this reason it has a form depending on boundaries. In particular, the definition based on the Fourier transform can not be applied to the fractional Laplacian acting in a bounded medium.

The statement of such a problem should be accompanied with a specification of the desired function values throughout an outer region. So, we have to return to the integral representation of the operator. The random flight interpretation can help in specifying the conditions but some subtle points such as distinction between first-passage and first-arrival times or between free and reflecting boundary conditions appear \cite{Zoia2007}. In \cite{Zoia2007} have investigated the matrix representation of the one-dimensional fractional Laplacian and solved numerically in connection to the first-passage problem (the L\`{e}vy-flights under absorbing boundary conditions) and to the long-ranged interfaces with no constraints at the ends (the free boundary conditions).

Krepysheva et al (\cite{Krepysheva2006}) analyze the symmetric
L\`{e}vy flights restricted to a semi-infinite domain by a
reflective barrier. They show that the introduction of the
boundary condition induces a modification in the kernel of the
nonlocal operator:
$$
-(-\triangle)^{\alpha/2}f(x,t)=
-\frac{1}{2\cos(\alpha\pi/2)\Gamma(2-\alpha)}\frac{\partial^2}{\partial x^2}\int\limits_0^\infty|x-\xi|^{1-\alpha}f(\xi,t)d\xi,\quad 1<\alpha<2,
$$
$$
\mapsto-(-\triangle)_{\mathrm {refl}}^{\alpha/2}f(x,t)=
-\frac{1}{2\cos(\alpha\pi/2)\Gamma(2-\alpha)}\frac{\partial^2}{\partial x^2}\int\limits_0^\infty\left[|x-\xi|^{1-\alpha}+(x+\xi)^{1-\alpha}\right]f(\xi,t)d\xi.
$$
The operators $-(-\triangle)^{\alpha/2}$ and
$-(-\triangle)_{\mathrm {refl}}^{\alpha/2}$ differ in the kernels,
but the difference becomes small when $x+\xi$ is large.
Nevertheless, omitting the term $(x+\xi)^{1-\alpha}$ we would get
a decreasing integral with respect to $x$, whereas the total
amount of the diffusing matter should be preserved.

Rafeiro and Samko (\cite{Rafeiro2005}) have introduced a version
of the fractional Laplacian for a bounded domain as a
generalization of the Marchaud formula for one-dimensional
fractional derivatives on an interval $(a,b),\ -\infty<a<b\leq
\infty$, to the multidimensional case of functions defined on a
region $G\subset \mathbb{R}^d$:
$$
\mathbb{D}_G^\alpha f(\mathbf{x})=C(\alpha)\left[a_G(\mathbf{x})f(\mathbf{x})+
\int\limits_G\frac{f(\mathbf{x})-f(\mathbf{y})}{|\mathbf{x}-\mathbf{y}|^{d+\alpha}}d\mathbf{y}\right],\quad
\quad \mathbf{x}\in G\subset \mathbb{R}^d,
$$
where $\alpha\in(0,1)$,
$$
C(\alpha)=\frac{\alpha 2^{\alpha-1}\Gamma[(d+\alpha)/2]}{\pi^{d/2}\Gamma(1-\alpha/2)}
$$
and
$$
a_G(\mathbf{x})=\int\limits_{\mathbb{R}^d\setminus G}\frac{d\mathbf{y}}
{|\mathbf{x}-\mathbf{y}|^{d+\alpha}}.
$$
In other words, this is the Riesz fractional derivative of the zero continuation of $f(\mathbf{x})$ from $G$ to the whole space $\mathbb{R}^d$.

Guan \& Ma \cite{Guan2005}, investigating the reflected symmetric $\alpha$-stable processes, gave the name \textit{regional fractional Laplacian} to the limit
$$
-(-\triangle)_{G}^{\alpha/2}f(\mathbf{x})\equiv\lim\limits_{\varepsilon\downarrow 0}C(\alpha)
\int\limits_{G,\ |\mathbf{x}-\mathbf{y}|>\varepsilon}\frac{f(\mathbf{x})-f(\mathbf{y})}
{|\mathbf{y}-\mathbf{x}|^{d+\alpha}}d\mathbf{y},
$$
provided it exists.

For more detail, the reader can be referred to the articles
\cite{Suarez1997, Hu2000, Bogdan2000, Song2003, Denisov2008,
Jeng2010}. Better understanding of the Laplacian in a bounded
domain can be achieved on the base of the non-local operator
theory \cite{Vazqueza, Gunzburger2010, Du2011}.

This short review is done in order to underline that in
contradistinction to classical case, the fractional Laplacian
$\triangle^{\alpha/2}$ change its form in a bounded domain and
cannot be determined by its Fourier transform
$-|\mathbf{k}|^{\alpha}$ anymore. Consequently, all results
obtained by Lagutin et all in 2001-2011 years relate to infinite
unbounded fractal medium. One should say, that referring to the
normal model with the use of Gaussian distribution in a bounded
model can not justify the similar use of the stable distributions
because their long tails may easily get the boundary surfaces
which are inaccessible for the normal Gaussian process.

\section{Numerical simulation}

\begin{figure}[thb]
\centering
\includegraphics[width=0.4\textwidth]{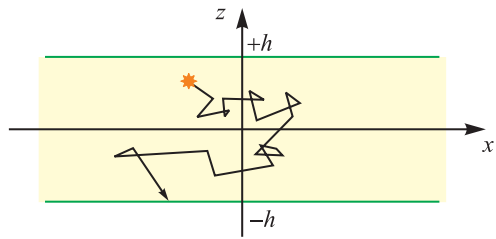}
\caption{The scheme of the model ($h=150$~pc).}\label{scheme}
\end{figure}

\begin{figure}[thb]
\centering
\includegraphics[width=0.9\textwidth]{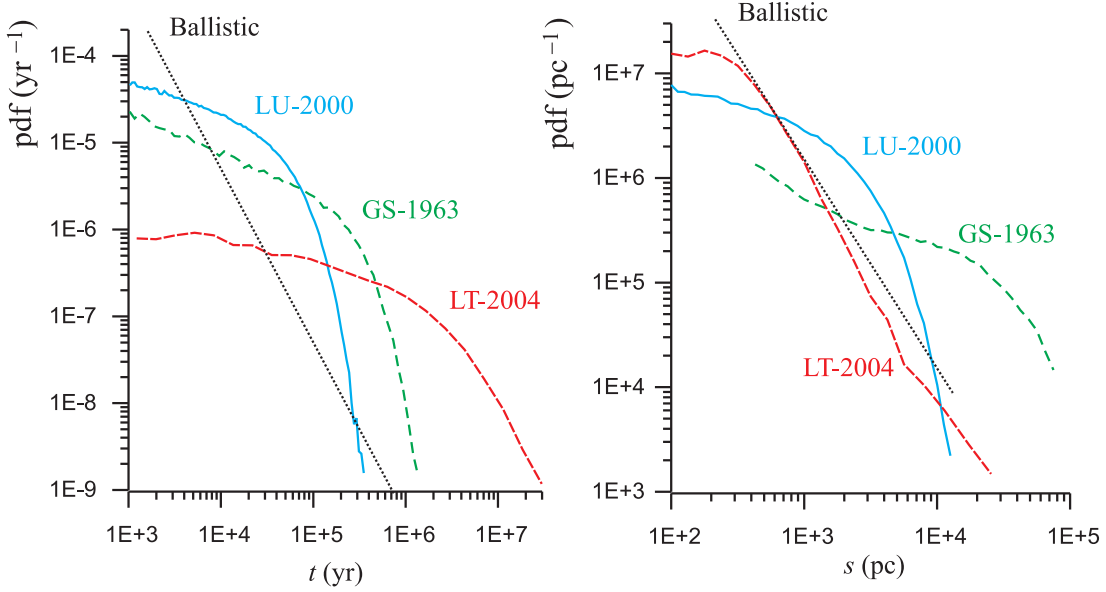}
\caption{Numerically calculated pdf of first passage time (left panel) and path (right panel) in frameworks of GS-1963,  LU-2000 and LT-2004 models.}\label{results}
\end{figure}

Taking into account the above-mentioned difficulties with
statement of boundary conditions  in analytic or numerical
approach, we perform direct Monte Carlo simulations to investigate
the CR propagation in the framework of a fractal Galaxy model.
The first problem we consider here is the escape time distribution
for the Galactic disk. As can be concluded from \cite{Taillet2003} (see also \cite{Berezinsky1990book}),
the leading contribution in this process belongs to plane boundaries so the escape through the cylindrical part of the boundary can in
the first approximation be neglected. Thus, we will simulate isotropic walk of particles in a fractal layer with two plane-parallel
boundaries and the initial random point uniformly distributed between these boundaries (Figure~\ref{scheme}). Objects for study are escape time and
escape path distributions in two models: LU ($\alpha=1.67,\quad \beta=1$, $v=c$) and LT ($\alpha=0.3$, $\beta=0.8$, $v=\infty$).
Results of Monte Carlo simulation are presented in Figure~\ref{results}.  The coefficient of anomalous diffusion
for both models $D=D_0 R^{0.27}$, $E=10^6$~GeV, parameter
${D_0=4\cdot10^{-6}}$~$\mathrm{pc}^{0.3}/\mathrm{year}^{0.8}$ in
LT-model and
${D_0=2.4\cdot10^{-3}}$~$\mathrm{pc}^{\alpha}/\mathrm{year}^{0.8}$ in
LU-model.
In the LT-model, distributions of free path lengths and waiting times have the form of asymptotical power laws
$$
{\rm P} \{\xi > r\} \sim \frac{(c_\alpha r)^{-\alpha}}{\Gamma(1-\alpha)},\quad \alpha > 0, \quad r
\to \infty; \quad {\rm P} \{\tau > t\} \sim\frac{(c_\beta t)^{-\beta}}{\Gamma(1-\beta)},\quad \beta > 0 \quad t \to
\infty.
$$
We take $c_\alpha=270$~pc$^{-1}$ and $c_\beta=10^{-2}$~yr$^{-1}$ (for $E=10^6$~GeV) and simulate $\xi$ and $\tau$ as random variables with pdf in the form of fractional exponents:
$$
\xi=\frac{|\ln U|^{1/\alpha}}{c_\alpha^{1/\alpha}}S(\alpha),\quad \tau=\frac{|\ln U|^{1/\alpha}}{c_\beta^{1/\alpha}}S(\beta).
$$
Here $S(\alpha)$ and $S(\beta)$ are one-sided stable variables simulated according to Kanter's algorithm
$$
S(\alpha)\stackrel{d}{=}
\frac{\sin(\alpha\pi U_2)[ \sin((1-\alpha)\pi
U_2)]^{1/\alpha-1}}{[\sin (\pi U_2)]^{1/\alpha}[\ln U_3]^{1/\alpha-1}},
$$
where $U_1$, $U_2$ and $U_3$ are variables uniformly distributed in $(0,1]$.

In the LU-model, waiting times can be simulated according to the exponential distribution
$$
{\rm P} \{\tau > t\} =\exp(-\mu t).
$$
We take $\mu=10^{-2}$~yr$^{-1}$. Path length are simulated according to the Pareto distribution ${{\rm P} \{\xi > r\}=b\ r^{-\alpha}}$
with $b=8\cdot10^{-2}$~pc$^{-\alpha}$. These parameters are valid for $E=10^6$~GeV.

Black dotted line corresponds to the ballistic motion
from a source. Distance between planes is equal to $2h=300$~pc. Right panel of Figure~3 shows the numerically calculated pdf of first passage time. Blue solid lines are for
LU-model and red dashed line is for LU model for the same
parameters as in the left panel.

\begin{figure}[thb]
\centering
\includegraphics[width=0.7\textwidth]{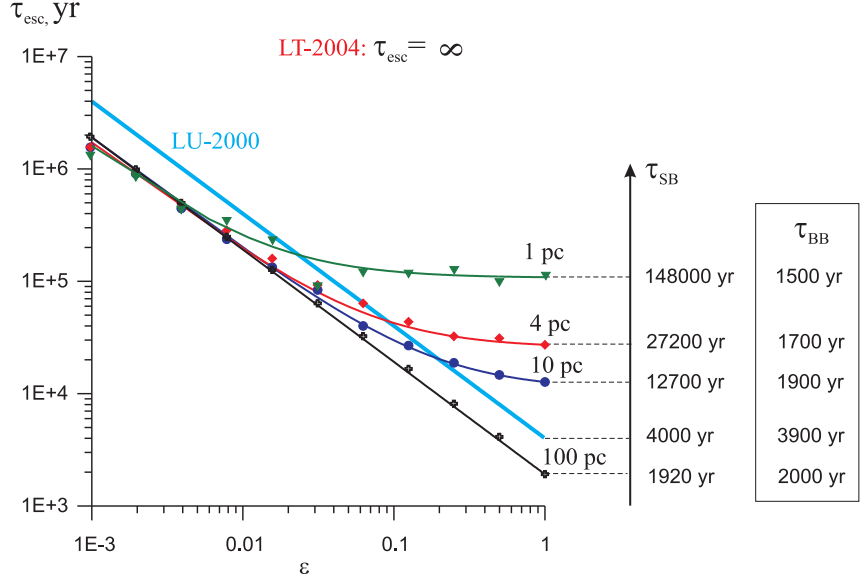}
\caption{Escape time versus transparency. The instantaneous point source is situated on the middle plane.}\label{escape}
\end{figure}

\begin{figure}[thb]
\centering
\includegraphics[width=1\textwidth]{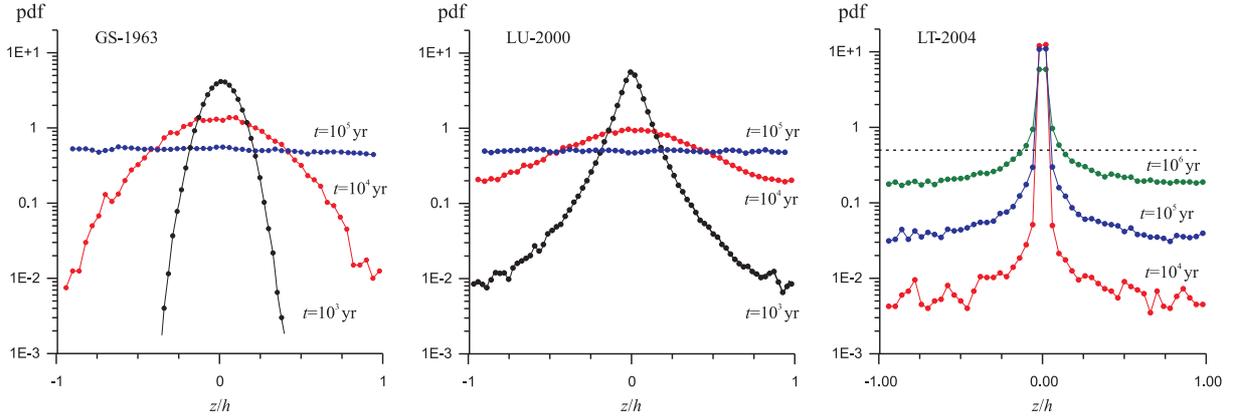}
\caption{Distribution density of the transverse coordinate $z$ of the cosmic rays in the region with specularly reflecting boundaries $z=\pm h$. The instantaneous point source is on the middle plane.}\label{equi}
\end{figure}

The mean escape time can be calculating according to the following formula
$$
\tau_{\mathrm{esc}}=\varepsilon\tau_{SB}+\varepsilon(1-\varepsilon)[\tau_{SB}+\tau_{BB}]
+\varepsilon(1-\varepsilon)^2[\tau_{SB}+2\tau_{BB}]+\dots
=\tau_{SB}+\frac{1-\varepsilon}{\varepsilon}\tau_{BB},
$$
where $\varepsilon$ is the transparency of boundaries, $\tau_{SB}$ is the mean passage time from source to boundary, and $\tau_{SB}$ is the mean passage time from boundary to boundary. In Figure~\ref{escape}, the dependences of escape time on transparency in the models under consideration are shown. Corresponding values of $\tau_{SB}$ and $\tau_{BB}$ are indicated in the figure. In the LT-2004 model the mean escape time is infinite due to trapping times with asymptotically power law distributions.

In our work \cite{Uchaikin2011b}, it has been shown by Monte
Carlo simulation that the LT-model provides large anisotropy for cosmic rays propagated in infinite space from a single source.
From Figure~\ref{results}, we can see that even specularly reflecting boundaries can not change this situation. First passage time are distributed in very wide interval. Figure~\ref{equi} confirms this reasoning. It shows
distribution density of the transverse coordinate $z$ for the case of random walk of a particle between two specularly reflecting boundaries with coordinates $z=\pm h$. Random walk starts from the middle plane $z=0$. The densities are calculated for several times. One can see that even for $t=10^6$~yr stationary distribution is not established. For the LU-model, the uniform distribution of transverse coordinate takes place at time $t\approx5\cdot10^4$~yr for parameters indicated above.

\section{Acknowledgments}
We thank the Russian Foundation for Basic Research (grants 10-01-00608, 11-01-00747, 12-01-00660) and the Ministry of Education and Science of the Russian Federation (grant 2.1894.2001) for financial support.

\end{document}